\documentstyle[12pt,preprint,aps]{revtex}

\begin{document}

\title{Parallel flow in Hele-Shaw cells with ferrofluids}

\author{Jos\'e A. Miranda}
\address{Laborat\'{o}rio de F\'{\i}sica Te\'{o}rica e Computacional,
Departamento de F\'{\i}sica,\\ Universidade Federal de Pernambuco
Recife, PE  50670-901 Brazil}

\author{Michael Widom}
\address{Department of Physics, Carnegie Mellon University, 
Pittsburgh, PA  15213 USA}
\date{\today}
\maketitle

\begin{abstract}
Parallel flow in a Hele-Shaw cell
occurs when two immiscible liquids flow with relative velocity
parallel to the interface between them. The interface is unstable due
to a Kelvin-Helmholtz type of instability in which fluid flow couples
with inertial effects to cause an initial small perturbation to
grow. Large amplitude disturbances form stable solitons. We consider
the effects of applied magnetic fields when one of the two fluids is a
ferrofluid. 
The dispersion relation governing mode growth is modified so that the
magnetic field can destabilize the interface even in the absence of
inertial effects. However, the magnetic field does not affect the speed
of wave propogation for a given wavenumber. We note that the magnetic
field creates an effective interaction between the solitons.
\end{abstract}


The Saffman-Taylor problem~\cite{Saf}
considers two immiscible viscous fluids moving in a narrow space
between two parallel plates (the so-called Hele-Shaw cell). When a low
viscosity fluid invades a region filled with high viscosity fluid, the
initially flat fluid-fluid interface is unstable and evolves through a
mechanism known as viscous fingering~\cite{Rev}. We call the
displacement of one fluid by another {\it frontal flow}. In contrast,
{\it parallel flow} occurs when the fluids flow parallel to the
interface separating them.  One important example of parallel flow
occurs after the passage of a fully developed Saffman-Taylor finger.

Recent experimental and theoretical studies~\cite{Zey1,Zey2,Gon1}
examined the dynamics of fluid interfaces under parallel flow in
Hele-Shaw cells.  Zeybek and Yortsos~\cite{Zey1,Zey2} studied, both
theoretically and experimentally, parallel flow in a horizontal
Hele-Shaw cell in the large capillary number limit.  For finite
capillary number and wavelength, linear stability analysis indicates
that small perturbations decay, but the rate of decay vanished in the limit
of large capillary numbers and large wavelength.  Furthermore, a
weakly nonlinear analysis of the problem found Korteweg-de Vries
(KdV) dynamics leading to stable finite amplitude soliton solutions.
Solitons were indeed observed experimentally.  Gondret and
Rabaud~\cite{Gon1} incorporated inertial terms into the equation of
motion in a Hele-Shaw cell and found a Kelvin-Helmholtz instability
for inviscid fluids.  For viscous fluids they derived a
Kelvin-Helmholtz-Darcy equation and found the threshold for
instability was governed by inertial effects, while the wave velocity
was governed by the Darcy's law flow of viscous fluids. Their
experimental results supported their theoretical analysis.

As was the case for frontal flow of nonmagnetic fluids in Hele-Shaw
cells, many research groups have studied the frontal interface
behavior when one of the fluids is a ferrofluid~\cite{Ros}, and an
external magnetic field is applied~\cite{Ros,Tse,WM,Fla,Za}. Ferrofluids,
which are colloidal suspensions of microscopic permanent magnets,
respond paramagnetically to applied fields. As a result of the
ferrofluid interaction with the external field, the usual frontal
displacement viscous fingering instability is supplemented by a
magnetic fluid instability~\cite{Ros}, resulting in a variety of new
interfacial behaviors. Depending on the applied field direction, one
observes highly branched, labyrinthine structures~\cite{Tse,WM},
patterns showing an ordered line of peaks~\cite{Fla}, or even the
supression of the usual viscous fingering instability~\cite{Za}.
Rosensweig~\cite{Ros} discusses the Kelvin-Helmholtz instability for
unconfined ferrofluids.

In this paper we perform the linear stability analysis for parallel
flow in which one fluid is a ferrofluid and a magnetic field is
applied. We consider three separate field configurations: (a) {\it
tangential}, for in-plane fields tangent to the unperturbed interface;
(b) {\it normal}, for in-plane applied fields normal to the
unperturbed interface; (c) {\it perpendicular}, when the field is
perpendicular to the plane defined by the Hele-Shaw cell plates. We
show the magnetic field provides additional mechanisms for
destabilizing the interface, and we analyze qualitatively the
interactions between solitons caused by the magnetic field. We
neglect inertial terms because they are not needed to understand the
interfacial instability.

Let us briefly describe the physical system of interest. Consider two
semi-infinite immiscible viscous fluids, flowing with velocities
$U_{1}$ and $U_{2}$, along the $x$ direction, in a Hele-Shaw cell of
thickness $b$ (see figure 1). We assume that $b$ is smaller than any
other length scale in the problem, and therefore the system is
considered to be effectively two-dimensional. Denote the densities and
viscosities of the lower and upper fluids, respectively as $\rho_{1}$,
$\eta_{1}$ and $\rho_{2}$, $\eta_{2}$. To achieve steady-state
parallel flow the velocities and viscosities must obey the condition
$\eta_1 U_1 = \eta_2 U_2$.
According to Gondret and Rabaud~\cite{Gon1}, we may neglect inertial
terms relative to viscous terms provided $kb << 12/{\rm Re}$, where
$k$ is a typical wavevector and ${\rm Re}=\rho U b/\eta$ is a
characteristic Reynold's number.

Between the two fluids there exists a surface tension $\sigma$. We
assume that the lower fluid is the ferrofluid (magnetization $\vec
M$), while the upper fluid is nonmagnetic. In order to include the
acceleration of gravity $\vec g$, we tilt the cell so that the $y$
axis lies at angle $\beta$ from the vertical direction.  To include
magnetic forces, we apply a uniform magnetic field $\vec H_0$, which
may point along the $x$, $y$ or $z$ axis. During the flow, the
fluid-fluid interface has a perturbed shape described as $y
=\zeta(x,t)$ (solid curve in figure 1).

Hydrodynamics of ferrofluids departs from the usual Navier-Stokes
equations through the inclusion of a term representing magnetic force.
Let ${\vec M}$ represent the local magnetization of the ferrofluid,
and note that the force on ${\vec M}$ depends on the gradient of local
magnetic field ${\vec H}$. The local field differs from the applied
field $\vec H_0$ by the demagnetizing field of the polarized
ferrofluid. We will assume ${\vec M}$ takes a constant value parallel
to the applied field. This amounts to neglecting the demagnetizing
field relative to the applied field and can be justified for low
magnetic susceptibility of the ferrofluid, or for large applied fields
that saturate the ferrofluid magnetization. It can also be justified
for very thin ferrofluid films when the field is parallel to the plane
of the cell.

For the quasi two-dimensional geometry of a Hele-Shaw cell, the three
dimensional flow may be replaced with an equivalent two-dimensional
flow ${\vec v}(x,y)$ by averaging over the $z$ direction perpendicular
to the plane of the Hele-Shaw cell. Imposing no-slip boundary
conditions and a parabolic velocity profile one derives Darcy's law
for ferrofluids in a Hele-Shaw cell~\cite{Jac,Tse2},
\begin{equation}
\label{Darcy}
\eta \vec v= -\frac{b^{2}}{12} \left \{ \vec \nabla p -
\frac{1}{b} \int_{-b/2}^{+b/2} (\vec M \cdot \vec \nabla) \vec H dz
- \rho (\vec g \cdot \hat{y})\hat{y} \right \},
\end{equation}
where $p$ is the hydrodynamic pressure. Equation~(\ref{Darcy}) 
describes nonmagnetic fluids by simply dropping
the terms involving magnetization.

When the velocity field ${\vec v}$ is irrotational, it is convenient
to rewrite equation~(\ref{Darcy}) in terms of velocity potentials. We
write $\vec v=-\vec\nabla\phi$, where $\phi$ defines the velocity
potential.  Likewise we introduce the scalar magnetic potential 
\begin{equation}
\label{magpot}
\varphi=
\int_{{\cal S}} \frac{\vec M \cdot \vec n'}{| \vec r - \vec r'|} d^{2}r'
\end{equation}
where $\vec H=- \vec \nabla \varphi$. Here the unprimed coordinates
$\vec r$ denote arbitrary points in space. The primed coordinates
$\vec r'$ are integration variables within the magnetic domain ${\cal
S}$, and $d^{2}r'$ denotes the infinitesimal area element. The vector
$\vec n'$ represents the unit normal to the magnetic domain in
consideration.

To study the interface dynamics, we evaluate
equation~(\ref{Darcy}) for each of the fluids on the interface,
subtract the resulting equations from each other, and divide by the
sum of the two fluids' viscosities to get the equation of motion
\begin{eqnarray}
\label{difference}
& A & \left ( {{\phi_2 + \phi_1}\over{2}} \right ) +
\left ( \frac{\phi_2 - \phi_1}{2} \right ) = {{b^{2}}\over{12(\eta_{1} + \eta_{2})}} \times  \nonumber \\
&   & \Bigg \{ \sigma \kappa + \frac{1}{b}  \int_{-b/2}^{+b/2} (\vec M \cdot \vec  \nabla\varphi) dz +  (\rho_{2} - \rho_{1}) g \cos \beta~y \Bigg \}. \nonumber \\
\end{eqnarray}
To obtain~(\ref{difference}) we have used the pressure boundary
condition $p_{2} - p_{1}=\sigma\kappa$ at the interface, where
$\kappa=( \partial^2 \zeta/ \partial x^{2} ) [ 1 + ( \partial\zeta/
\partial x)^2]^{-3/2}$ denotes the interfacial curvature in the plane
of the Hele-Shaw cell. The dimensionless parameter $A=(\eta_{2} -
\eta_{1})/(\eta_{2} + \eta_{1})$ is the viscosity contrast.

We perturb the interface with a single Fourier mode
\begin{equation}
\label{expansion}
\zeta(x,t)=\zeta_{k} \exp(i(\omega t - k x)).
\end{equation}
The velocity potential for fluid $i$, $\phi_{i}$, must contain the
uniform unperturbed flow $U_i$ and a perturbed part that reflects the
space and time dependence of $\zeta$, obeys Laplace's equation
$\nabla^{2}\phi_{i}=0$ and vanishes as $y \rightarrow \pm \infty$.
The velocity potentials obeying these requirements are
\begin{equation}
\label{phi1-2}
\phi_{i}=\phi_{i k} \exp(\pm|k|y) \exp(i(\omega t - k x)) - U_{i}~x.
\end{equation}

To conclude our derivation and close equation~(\ref{difference}) we
need additional relations expressing the velocity potentials in terms
of the perturbation amplitudes. To find these, we considered the kinematic
boundary condition, which states that the normal components of each
fluid's velocity at the interface equals the normal velocity of the
interface itself~\cite{Ros,WM}.
Inserting expression~(\ref{expansion}) for $\zeta(x,t)$
and~(\ref{phi1-2}) for $\phi_1$ into the kinematic boundary condition,
we solved for $\phi_{ik}(t)$ consistently to
first order in $\zeta$ to find
\begin{equation}
\label{phi1t}
\phi_{1k} = -\frac{i \omega \zeta_k }{|k|} 
+ i \frac{k}{|k|} ~U_{1} \zeta_{k},
\end{equation}
and a similar expression for $\phi_{2k}$.

Substitute expression~(\ref{phi1t}) for $\phi_{1k}$ and the related
expression for $\phi_{2k}$ into equation of motion~(\ref{difference}),
and again keep only linear terms in the perturbation amplitude. 
This procedure eliminates the velocity potentials from
equation~(\ref{difference}), and we obtain the dispersion relation for
growth of the perturbation $\zeta(x,t)$
\begin{equation}
\label{dispersion}
\omega=k \left ( \frac{\eta_{1}U_{1} + \eta_{2}U_{2}}{\eta_{1} + \eta_{2}} \right ) - \frac{i |k|\sigma}{12(\eta_{1} + \eta_{2})} \left [ N_{B} I_{j}(k) - (kb)^{2} - (k_{0}b)^{2} \right ] 
\end{equation}
where
$N_{B}=2M^{2}b/\sigma$ is the magnetic Bond number and
$k_{0}=\sqrt{[(\rho_{1} - \rho_{2})g \cos \beta]/\sigma}$.

The real part of $\omega$ is $k$ times the phase velocity, and is the
viscosity-weighted average of the two fluid velocities. Note that the
magnetic field does not alter the phase velocity of the waves.  The
imaginary part of $\omega$, which governs the exponential growth or
decay of the wave amplitude, does include effects of the magnetic
field. Exponential (unstable) growth occurs when the imaginary part of
$\omega$ is negative.  We point out that when there is no applied
magnetic field $(N_{B}=0)$ our equation~(\ref{dispersion}) agrees with
the dispersion relation derived by Gondret and Rabaud~\cite{Gon1} for
the case in which the cell is vertical ($\beta=0$) and by Zeybek and
Yortsos~\cite{Zey1,Zey2} for the case in which the cell is horizontal
($\beta=\pi/2$).

Terms containing $I_{j}(k)$ originate from the Fourier transforms of 
\begin{equation}
\label{Ijx}
M^2 I_j(x) \equiv {{1}\over{b}}
\int_{-b/2}^{+b/2}  M_j \frac{\partial\varphi}{\partial r_j} ~dz,
\end{equation}
the magnetic contribution to equation~(\ref{difference}). The
subscript $j=x, y, z$ indicates the tangential, normal and
perpendicular magnetic field configurations, respectively.
For $\vec{M}$ in the $x$ or $y$ direction we can expand equation~(\ref{Ijx}) 
to first order in $\zeta$ to obtain
\begin{equation}
\label{Ix}
I_x(x)=\int_{-\infty}^{\infty} dx' (x-x') \left [ -\frac{\partial \zeta(x')}{\partial x'} \right ] \bar{F}(x-x')
\end{equation}
and
\begin{equation}
\label{Iy}
I_y(x)=\int_{-\infty}^{\infty} dx' [\zeta(x)-\zeta(x')] \bar{F}(x-x')
\end{equation}
where
\begin{equation}
\bar{F}(x) \equiv \frac{1}{b} \int_{-b/2}^{+b/2}\int_{-b/2}^{+b/2} 
\frac{dz dz'}{[x^2+(z-z')^2]^{3/2}}\\
={{2}\over{b x^2}} [\sqrt{b^2+x^2}-|x|].
\end{equation}
In contrast, for $I_z(x)$ the $z$ integration inverts the derivative
of $\varphi$ with respect to $z$ in equation~(\ref{Ijx}) so that 
after integrating over $y'$ and expanding to first order in
powers of $\zeta$, this term simplifies to
\begin{equation}
\label{Iz2}
I_z(x)= \int_{-\infty}^{\infty} dx' {{2}\over{b}}
\left [ \frac{1}{\sqrt{(x-x')^2}}-\frac{1}{\sqrt{(x-x')^2+b^2}} \right ] 
[\zeta(x')-\zeta(x)].
\end{equation}

We obtain the specific forms for the $I_{j}(k)$'s corresponding to
each particular field configuration by taking the Fourier transform of
equations~(\ref{Ix}),~(\ref{Iy}), and~(\ref{Iz2}). After some simple
algebra we find the following expressions for the magnetic terms
$I_{j}(k)$

\begin{equation}
\label{Ikt}
I_{x}(k)= - 2~\int_{0}^{\infty} \left ( \frac{\sin \tau}{\tau} \right ) [\sqrt{(kb)^{2} + \tau^{2}} - \tau]~d \tau,
\end{equation}

\begin{equation}
\label{Ikn}
I_{y}(k)=4~\int_{0}^{\infty} \left ( \frac{\sin \tau}{\tau} \right )^{2} 
[\sqrt{(kb/2)^{2} + \tau^{2}} - \tau]~d \tau,
\end{equation}

and

\begin{equation}
\label{Ikp}
I_{z}(k)=4 \int_{0}^{\infty}  \sin^{2} \tau \left [ \frac{1}{\tau}-\frac{1}{\sqrt{(kb/2)^2+\tau^2}} \right ]~d \tau.
\end{equation}

In the limits of small and large wavevector these Fourier transforms
reduce to
\begin{equation}
\label{Ixlim}
I_x(k) \approx \left\{ \begin{array}{ll}
			-[(3/2 - C + \ln{2})-\ln{k b}](k b)^2&	~~~kb << 1\\
			-\pi kb			&	~~~kb >> 1
			\end{array} \right.
\end{equation}
\begin{equation}
\label{Iylim}
I_y(k) \approx \left\{ \begin{array}{ll}
			[(2 - C + \ln{2})-\ln{k b}](kb)^2/2&	~~~kb << 1\\
			\pi kb			&	~~~kb >> 1
			\end{array} \right.
\end{equation}
\begin{equation}
\label{Izlim}
I_z(k) \approx \left\{ \begin{array}{ll}
			[(1 - C + \ln{2})-\ln{kb}](kb)^2/2&	~~~kb << 1\\
			\ln{(kb/2)}			&	~~~kb >> 1,
			\end{array} \right.
\end{equation}
where $C \approx 0.57721$ denotes Euler's constant~\cite{Mag}. 
Our results~(\ref{Ikt}), ~(\ref{Ikn}) and~(\ref{Ikp}) agree with similar
kind of calculations related to {\it frontal} displacements in Hele-Shaw
cell with ferrofluids~\cite{Tse,WM,Fla,Za}.

The dispersion relation~(\ref{dispersion}) is given for the case of
systems with infinite extent along the $y$-axis. For finite extent $L$
the algebraic dependence on wavevector $k$ is modified by a first order
rational function of $\sinh{k L}$ as shown by Zeybeck and
Yortsos~\cite{Zey1,Zey2}. When $kL$ is large this finite size
correction dies off exponentially quickly. The magnetic integrals $I_j(k)$
likewise possess exponentially small finite size corrections.

Consider the stability of the fluid-fluid interface for the different
field configurations. The initially flat interface is unstable to
perturbations with wavenumber $k$ when $N_B I_j(k)-(k b)^2-(k_0 b)^2$
is positive.  If the heavier fluid is below the lighter fluid,
$(\rho_{1} > \rho_{2})$, then both gravity and surface tension
stabilize the system and $k_0$ is real. Therefore, in the absence of
applied magnetic field $(N_{B}=0)$, the temporal growth rate of any
perturbation is negative and waves are damped.  On the other hand, if
the external magnetic field is nonzero, the stability of the interface
will depend on the field's direction.  Figure 2 illustrates how the
magnetic terms~(\ref{Ikt}), ~(\ref{Ikn}) and~(\ref{Ikp}) vary with
reduced wave number $kb$. Inspecting figure 2 and the imaginary part
of the dispersion relation~(\ref{dispersion}) we note that a tangent
field configuration ($I_{x}(k) < 0$), makes the growth rate even more
negative than when the field is absent. So a tangent external field
has a stabilizing nature, reinforcing the effects of gravity and
surface tension. In contrast, since $I_{y}(k)$ and $I_{z}(k)$ are both
positive quantities, if a sufficiently strong magnetic field is
applied normal to the fluid-fluid interface, or perpendicular to the
cell plates, the growth rate may become positive, leading to a
possible destabilization of the interface.  We conclude that the
magnetic field can destabilize the interface even in the absence of
inertial effects.

In addition to the interface stability issue discussed above, it is
interesting to ask how the magnetic field acts on the motion of
interfacial waves once they appear. In the following, we discuss the
action of the applied magnetic field on the solitons that appear in
parallel flow in Hele-Shaw cells. To treat the problem rigorously
would require reproducing the analysis of Zeybek and
Yortsos~\cite{Zey1,Zey2} that derived Airy and KdV equations from a
weakly nonlinear analysis of the interfacial perturbations. Here we
simply point out that the solitons may be considered as localized
perturbations on the flat interface. When magnetic fields are present
the solitons acquire net dipole moments equal to the magnetization
of the fluid multiplied by the integrated area of the soliton.

Take the generic form of a KdV soliton,
\begin{equation}
\label{soliton}
u(x,t)=-{{c}\over{2}} {\rm sech}
\left ( {{\sqrt{c}}\over{2}} (x-ct) \right ),
\end{equation}
written here in terms of the scaled time, position and height variables
discussed in~\cite{Zey2}, where $c$ is the speed of propogation. 
We define the scaled dipole moment of the soliton of speed $c$ as
\begin{equation}
\label{moment}
m(c)=\int_{-\infty}^{\infty} {\vec M} u(x,t) dx 
= -\sqrt{c}\pi{\vec M}.
\end{equation}
In doing so, we neglect the magnetic field dependence of the shape of
the soliton. We may consider the magnetic moment~(\ref{moment}) as the
leading, linear term in a perturbative series in powers of applied
field, and expect a cubic correction due to the field-dependent
soliton shape. As noted in~\cite{Zey2}, the actual profile in unscaled
coordinates may be either positive or negative, and the dipole moment
given here must be divided by the position and height rescaling
factors to yield the true moment. True dipole moments $m$ point
parallel to the magnetization ${\vec M}$ when the soliton consists of
excess magnetic fluid, and points opposite to ${\vec M}$ when the soliton
consists of missing magnetic fluid.

Dipole interactions are long-ranged, falling off as $1/x^3$ for
moments separated by a distance $x$. This contrasts with the
fluid-dynamic interaction of solitons which decays exponentially with
separation. An interesting additional feature of the dipole-dipole
interaction is its variation with the relative orientation of dipole
moments and the vector joining them. In the case of solitons with parallel
moments $\vec{m}_1$ and $\vec{m}_2$ displaced from each other along
the $x$ axis, the interactions will be attracting, with strength
$2m_1m_2$, when the magnetizations lie along the $x$ axis (tangetial)
but will be repelling, with strength $m_1m_2$ when the magnetizations
lie along the $y$ (normal) or $z$ (perpendicular) axes.

In conclusion, we have performed the linear stability analysis for
parallel flow in a Hele-Shaw cell when one of the fluids is a
ferrofluid.  We show that the magnetic field may provide a new
mechanism for destabilizing the interface in the absence of inertial
effects, and we determine the magnetic correction to the dispersion
relations for three distinct field orientations. Finally, we suggest
parallel flow of ferrofluids as a novel system in which to
investigate soliton interactions.

\vspace{0.5 cm}
\begin{center}
{\bf ACKNOWLEDGMENTS}
\end {center}
\noindent
We thank J.C. Bacri for interesting discussions on the possibility of
conducting parallel flow experiments with ferrofluids.  J.A.M. would
like to thank CNPq (Brazilian Agency) for financial support. This work
was supported in part by the National Science Foundation grant
DMR-9732567.

\pagebreak
\noindent
\centerline{{\large {FIGURE CAPTIONS}}}
\vskip 0.5 in
\noindent
{FIG. 1:} Schematic configuration of the parallel flow geometry.

\vskip 0.25 in
\noindent
{FIG. 2:} Variation of $I_{j}(k)$ as a function of $kb$  for (a) tangential, 
(b) normal, and (c) perpendicular magnetic field configurations.

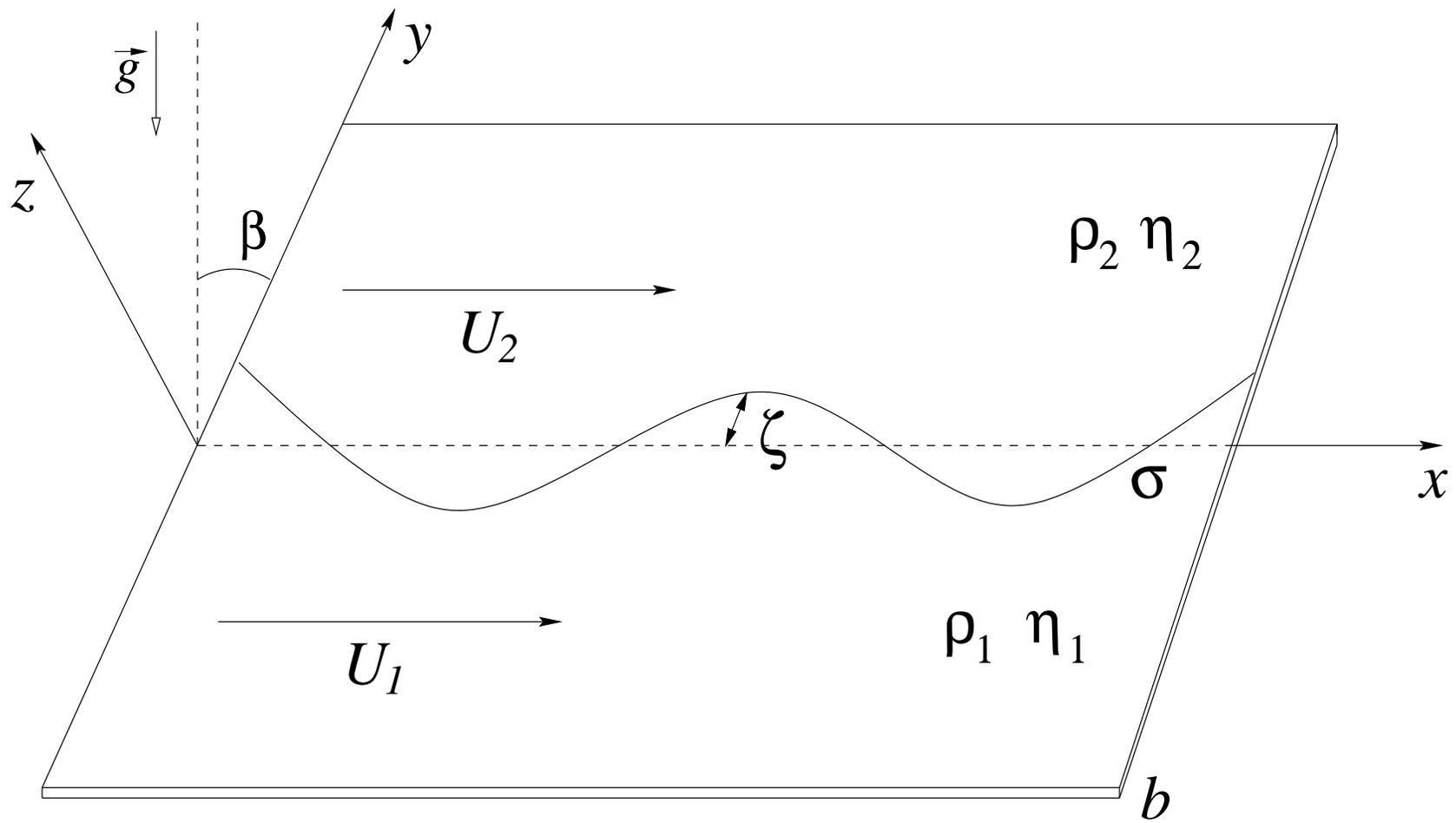
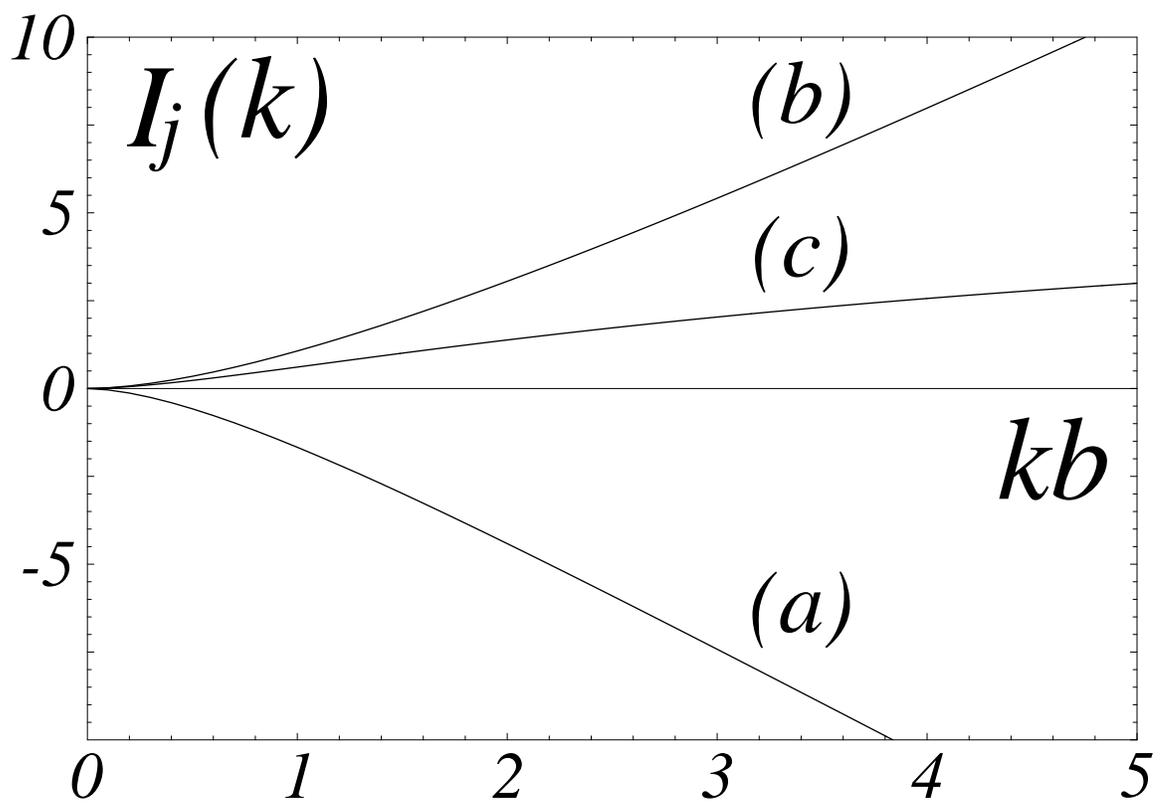
\end{document}